# Corotating Shock Waves and the Solar-Wind Source of Energetic Ion Abundances: Power Laws in *A/Q*

**Donald V. Reames**

Institute for Physical Science and Technology, University of Maryland, College Park, MD 20742-2431 USA, email: dvreames@umd.edu

**Abstract** We find that element abundances in energetic ions accelerated by shock waves formed at corotating interaction regions (CIRs) mirror the abundances of the solar wind modified by a decreasing power-law dependence on the mass-to-charge ratio *A/Q* of the ions. This behavior is similar in character to the well-known power-law dependence on *A/Q* of abundances in large gradual solar energetic particles (SEP). The CIR ions reflect the pattern of *A/Q*, with *Q* values of the source plasma temperature or freezing-in temperature of 1.0 –1.2 MK typical of the fast solar wind in this case. Thus the relative ion abundances in CIRs are of the form $(A/Q)^a$ where *a* is nearly always negative and evidently decreases with distance from the shocks, which usually begin beyond 1 AU. For one unusual historic CIR event where $a \approx 0$, the reverse shock wave of the CIR seems to occur at 1 AU, and these abundances of the energetic ions become a direct proxy for the abundances of the fast solar wind.

# 1. Introduction

Corotating interaction regions (CIRs) are regions of compressed plasma produced when high-speed solar-wind streams, emitted from solar coronal holes, collide with slow solar wind emitted earlier in the solar rotation (*e.g.* Richardson *et al.*, 1993, Mason *et al.*, 1997; Richardson, 2004; Bučík *et al.*, 2012). Two collisionless shock waves can emerge from these regions, a forward shock wave propagates outward into the slow solar wind (SSW) and a second, stronger, reverse shock wave propagates sunward into the fast solar wind (FSW). The shock waves usually begin to form outside the orbit of Earth and they may be strong enough to accelerate energetic ions from the solar wind. These CIR-associated ions propagate sunward from their source to take their place near Earth among the MeV ions from other sources, including impulsive (Mason, 2007; Reames, Cliver, and Kahler, 2014) or gradual (Kahler *et al.*, 1984; Lee, Mewaldt, and Giacalone, 2012; Desai and Giacalone, 2016) solar energetic-particle (SEP) events that have been extensively compared and reviewed (*e.g*. Gosling, 1993; Reames, 1999a, 2013, 2015, 2017a), the anomalous cosmic rays (ACRs; Reames, 1999b; Reames and McDonald, 2003) from the outer heliosphere, and the galactic cosmic rays (GCRs; Reames and McDonald, 2003) from beyond.

Different populations of energetic ions have different temporal behavior, are correlated with different solar phenomena, and may have different energy spectra or angular distributions. Over the years, however, we have found a primary signature of different populations of energetic ions to be the relative abundances of the chemical elements they include (*e.g.* Reames, 1999, 2017a). For example, it was recognized early that the average abundances in he large gradual SEP events have a distinctive dependence on the first ionization potential (FIP) of the elements (*e.g.* Webber, 1977; Meyer, 1985) where elements with FIP < 10 eV (*e.g.* Mg, Si, Fe) are enhanced by a factor of about 4 relative to those with FIP > 10 eV (*e.g.* He, O, Ne), when compared with corresponding abundances in the solar photosphere. Low-FIP elements are already ionized in the photosphere while high-FIP elements are neutral atoms; the ions may be preferentially boosted upward by the ponderomotive force of Alfvén waves (Laming, 2009, 2015), for example, as they expand across the chromosphere and into the high-temperature corona, where all elements become ionized.

This "FIP effect" is a property of the abundances of elements in the solar corona, relative to those in the photosphere. Therefore a modest FIP effect is also seen in the FSW and a larger one in the SSW (von Steiger *et al.*, 2000; Gloeckler and Geiss, 2007; Bochsler, 2009; Abbo *et al.*, 2016). However, recent measurements show that the FIP effect seen in SEPs is clearly different from that in the SSW or FSW (*e.g.* Mewaldt *et al.*, 2002; Reames, 2018a, 2018b). This difference has been recently understood (Reames, 2018a) in terms of the Alfvén-wave FIP model (Laming, 2009, 2015, 2017; Laming *et al.*, 2013) where material, that may eventually be shock accelerated to become SEPs, was convected into the corona on initially-closed magnetic-field lines in active regions, while that contributing to the solar wind is carried into the corona on open field lines. Alfvén waves can resonate with the field-line length on closed loops but not open on field lines, and model calculations are shown for each case by Laming (2015).

However, in addition to the average dependence of SEP abundances on FIP, as a consequence of their coronal origin, Meyer (1985) recognized that individual SEP events also had different dependences on *A/Q* of the ions. In fact, a power-law dependence on *A/Q* was dramatically shown by Breneman and Stone (1985) using newly-measured values of *Q* from Luhn *et al.* (1984). We have come to understand that this *A/Q* dependence comes from particle transport (Parker, 1963; Ng, Reames, and Tylka, 2003; Reames, 2016a, 2016b, 2018b) after acceleration by shock waves driven out by coronal mass ejections (CMEs). Observed at constant velocity, particle scattering during transport varies as a power of *A/Q* so that Fe scatters less than O, for example, causing Fe/O to be enhanced early in events and suppressed later; this dependence on time becomes a dependence on longitude because of solar rotation. It has recently been shown (Reames, 2016a, 2018b) that we may use of this power-law dependence to ask what source plasma temperature produces the most appropriate pattern of *A/Q* to yield a power law with an observed pattern of element enhancements relative to the average coronal abundances.

Impulsive SEP events also have a strong power-law dependence on *A/Q*. Once known only as a 10-fold enhancement in Fe/O, it has been extended to the 1000-fold enhancements of heavy elements like Au or Pb (Reames, 2000; Mason *et al.*, 2004; Reames and Ng, 2004, Reames, Cliver, and Kahler, 2014). These enhancements are ascribed to magnetic-rigidity-dependent enhancements during the collapse of reconnecting magnetic islands in solar jets and flares (Drake *et al.*, 2009). As a further complication, residual suprathermal ions from frequent small impulsive SEP events are sometimes incorporated into the seed population encountered by CME-driven shock waves where they may be preferentially re-accelerated to become gradual events (Reames, 2013, 2016b, 2017a).

If energetic particles from CIRs have a source in the FSW, we might expect that they are a proxy for FSW abundances and would provide an alternate measure of the results of the open field model of the FIP effect. Reames (2018a) used the CIR abundance measurements of Reames, Richardson, and Barbier (1991; see also Reames, 1995) and found that CIR events lent strong support for differences between SEP and solar wind abundances. However, not all measurements of energetic ion abundances in CIRs are the same. Richardson (2004) reviews abundance measurements where CIR abundances look more like those of SEPs at low solar-wind speeds, while Mason *et al.* (2008) and Bučík *et al.* (2012) find varying abundances in 40 or 50 CIR events at energies below about 1 MeV amu$^{-1}$. Perhaps some CIRs with low solar-wind speeds have incorporated SEPs directly or in their seed populations, but that cannot explain all the variation. Does the FSW itself vary or is the variation only in the accelerated ions? Apart from the possible reacceleration of SEPs, the observed variations in the abundances in CIRs remain uncertain. Can we understand these variations? Do CIR abundances have a power-law dependence on *A/Q* like those of SEPs? Can we derive a respectable estimate of FSW abundances from measurements of CIRs? The goal is to study CIRs that seen to accelerate the FSW, not those that may reaccelerate SEPs.

In this work we study element abundances in some of the largest CIR-associated events that occur during 24 years of observation by the Low-Energy Matrix Telescope (LEMT) on the *Wind* spacecraft, near Earth (von Rosenvinge *et al.*, 1995). For our purpose, LEMT primarily measures elements He through Fe in the region of 2 – 20 MeV amu$^{-1}$. The element resolution of LEMT is shown in detail by Reames (2014). We shall also have occasion to revisit the abundance measurements by the Very Low-Energy Telescope (VLET) on the International Sun-Earth Explorer (ISEE) 3. Element resolution of the VLET is shown by Reames (1995). The two telescopes are logically similar with similar element and energy coverage, although LEMT has a geometry factor of ≈ 50 cm$^2$ sr, while, unfortunately, that for VLET is below ≈ 1 cm$^2$ sr.

## 2. Selection of CIR Events

The selection of energetic particle increases in association with increases in solar-wind speed is relatively straightforward. Evidence of increasing density and magnetic-field strength for the formation of a reverse shock wave is helpful although none of our recent events actually have a shock formed at 1 AU. We need high intensities, often aided by 27-day recurrence, so as to measure as many elements between He and Fe as accurately as possible. We begin with the list of CIR events from Mason *et al.* (2008) and extend it to recent years, but events with ions below 1 MeV amu$^{-1}$ frequently do not have significant intensities above 1 MeV amu$^{-1}$ where we measure.

Before proceeding to study our events, it is important to show possible pitfalls and background sources that can intervene. Figure 1 shows a CIR event and three impulsive SEP events. Note that if the particles in the first SEP event and the CIR event were inadvertently combined, the SEP He, C, and O would make only a small contribution, but the SEP Fe contribution would dominate the mix. Since Fe/O ≈ 1 in impulsive SEP events, a small impulsive SEP background could pollute only the Fe and be otherwise invisible.

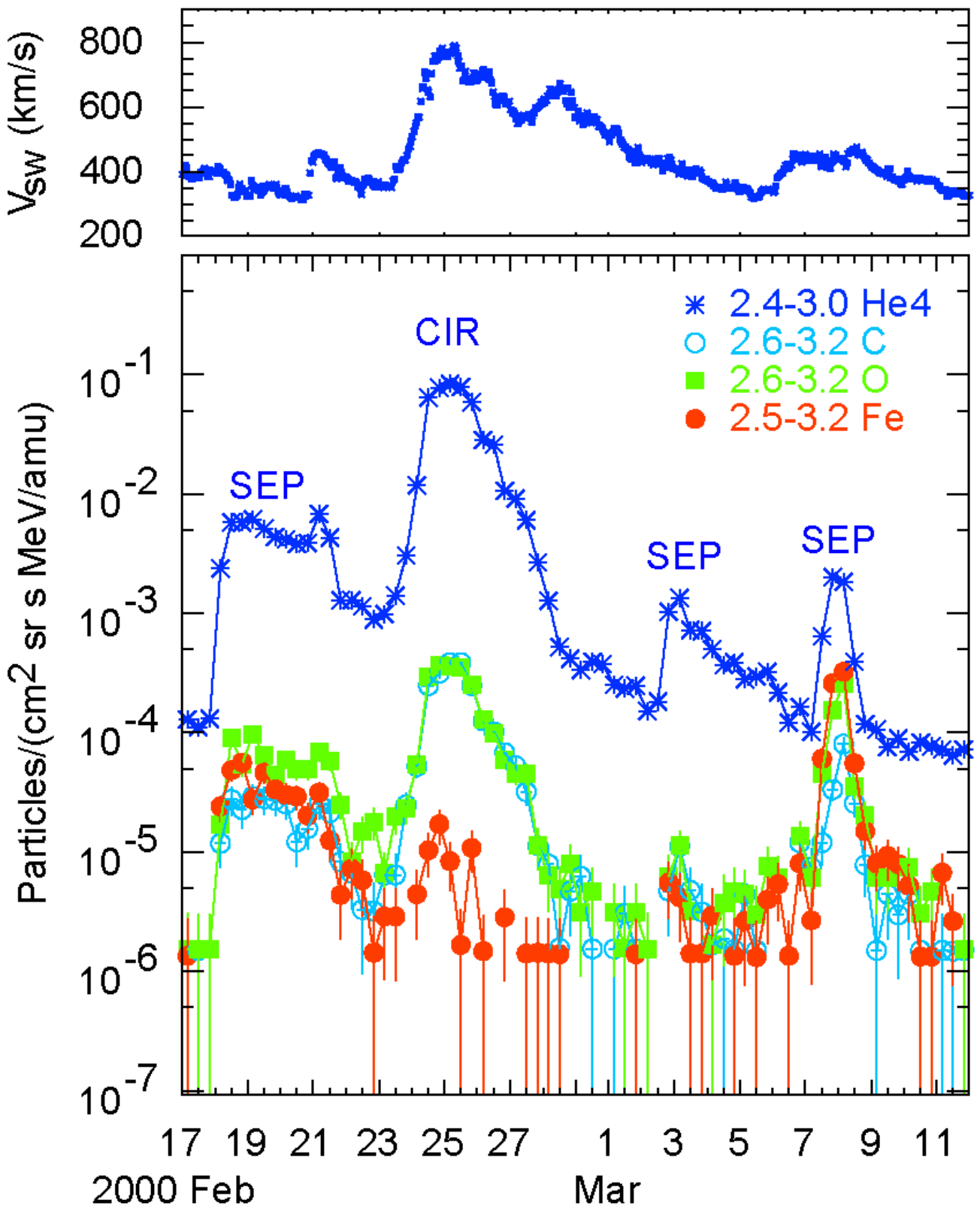

**Figure 1**. The *lower panel* shows intensities of $^4$He, C, O, and Fe at 2.5 – 3.2 MeV amu$^{-1}$ *vs.* time early in 2000. The CIR event and three impulsive SEP events are indicated. The SEP events have C/O < 0.5 and Fe/O ≈ 1, while the CIR event has C/O ≈ 1 and low Fe/O. The *upper panel* shows the solar-wind speed.

To avoid contamination from impulsive SEP events, one might be tempted to select CIRs away from solar maximum and nearer solar minimum. Then one might find the beautiful series of four recurrent CIR events as shown in Figure 2.

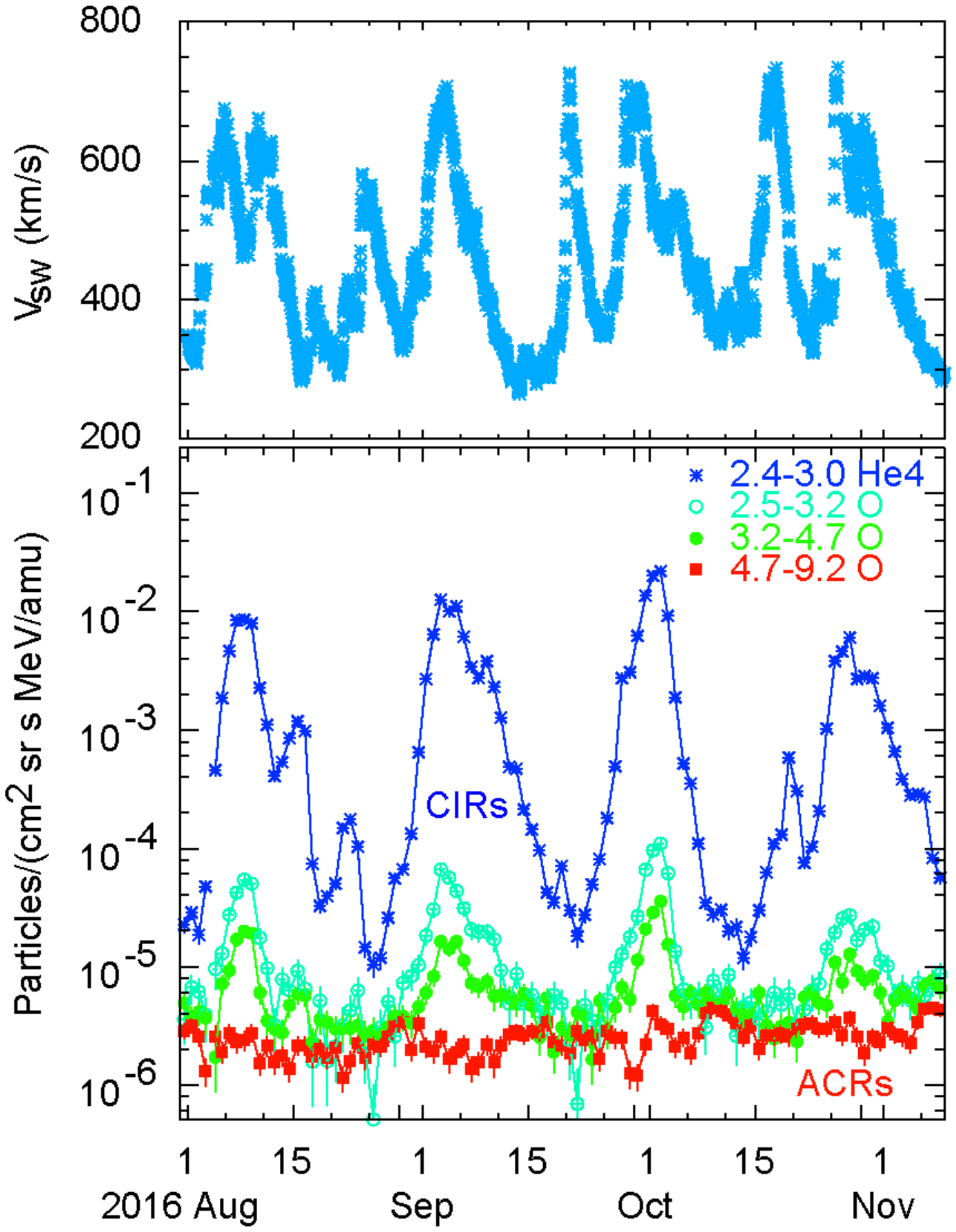

**Figure 2**. He and O ions, of the indicated energies, are shown in the *lower panel* while the solar-wind speed is shown in the *upper panel*. A series of four recurrent CIR events are flooded with O from ACRs at the higher energies.

The remarkable series of CIR events in Figure 2 is dominated by ACRs at the higher energies. At solar minimum O ≈ He ≈ $10^{-5}$ particles (cm$^2$ sr s MeV amu$^{-1}$)$^{-1}$ for ACRs, the spectra are flat down to below1 MeV amu$^{-1}$, and the abundances of other high-FIP ions, N, Ne, and Ar are also affected (*e.g.* Reames, 1999b, Reames and McDonald, 2003). Nevertheless, despite the ACRs, C/O ≈ 1 in the 2.5 – 3.2 MeV amu$^{-1}$ interval in these events. The presence of ACRs is usually obvious but it does complicate or invalidate otherwise interesting CIR events.

The CIR time intervals studied are listed in Table 1 along with other derived properties that will be derived subsequently.

# 3. CIR-Ion Energy Spectra

Figure 3 shows energy spectra for the first four of the study events. The lower panel compares C and O spectra normalized to the corresponding He spectra. Normalization factors will be considered in the next section. The theory of Fisk and Lee (1980) describes these spectra as a power law times an exponential in particle velocity, so the spectral shape in MeV amu$^{-1}$ should be the same for all ions. That seems to be the case for He, C, and O.

**Figure 3.** Energy spectra of the first four CIR events from Table 1 compare C and O with He in the *lower panel* and Fe with He in the *upper panel*. Spectral shapes for He, C, and O (and other species) generally agree well, but Fe shows possible background from impulsive SEPs at low energies in event 3.

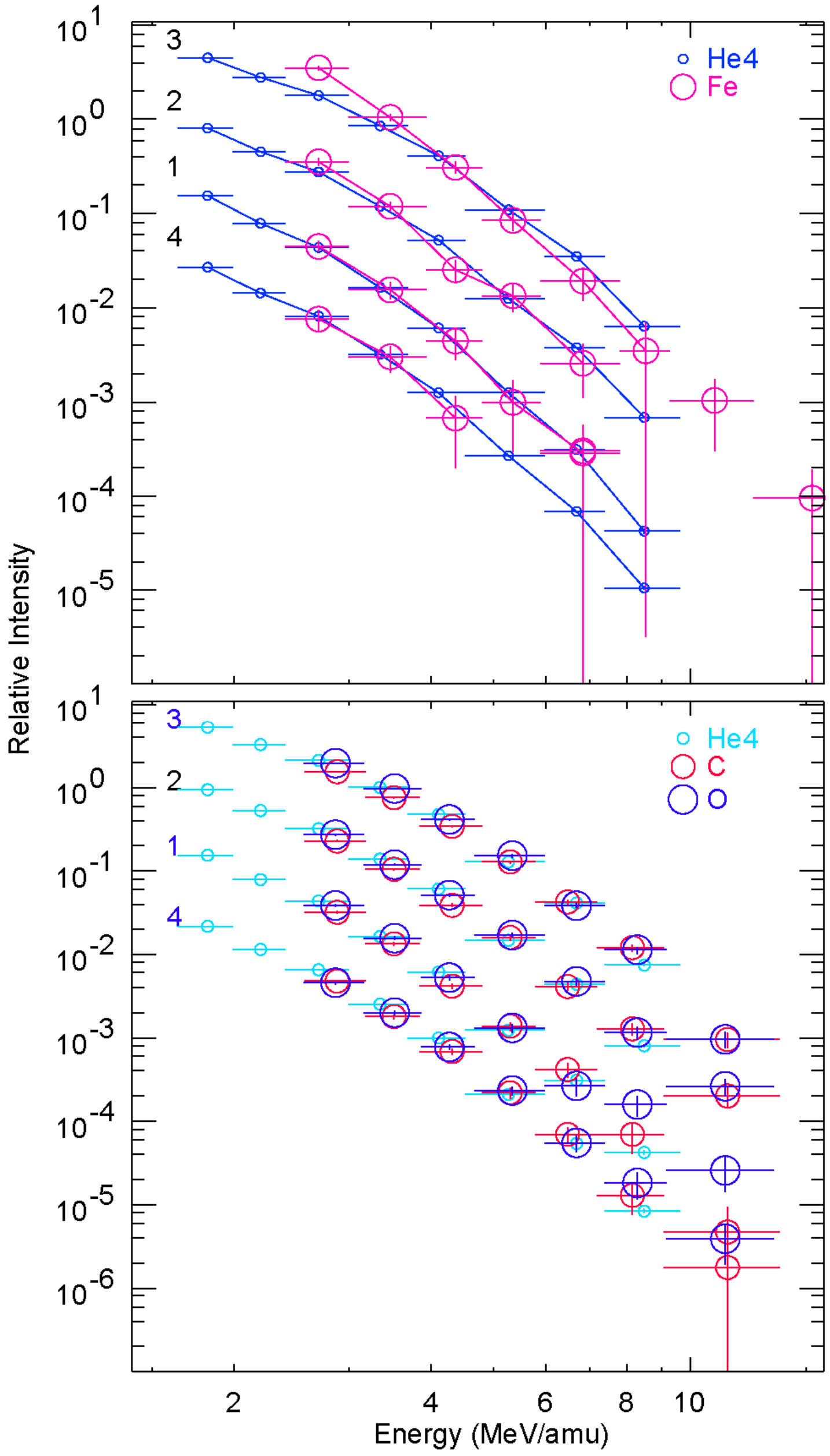

The upper panel of Figure 3 shows reasonable agreement of the shapes of He and Fe spectra for events 1, 2, and 4, but event 3 suggests the possible presence of excess low-energy Fe that may be background from an impulsive SEP event. The remaining CIR events in Table 1 have reasonably well-behaved spectra for all species. Event 3 is the greatest exception.

## 4. Relative Abundances of Elements

The normalization factors used to compare the spectra are the observed element abundances in each CIR event. To test for a dependence on $A/Q$, we can divide these abundances by those of the FSW (Bochsler, 2009) and plot them *vs.* $A/Q$. Values of $Q$ for the elements of the FSW correspond to a measured freezing-in temperature of about 1.1 MK (*e.g.* von Steiger *et al.*, 2000). Rather than using this value for all events, we search for the temperature $T$ that gives best fit power law in each event as has been done for gradual SEP events (Reames, 2016a, 2018b). Values of $\chi^2$ of the power-law fit divided by the number of degrees of freedom $m$ are shown *vs.* $T$ for all 12 CIR events in Figure 4 and the fits for all of the events are shown in Figure 5. The $T$ values of the minima seem to confirm the measured value of 1.1 MK.

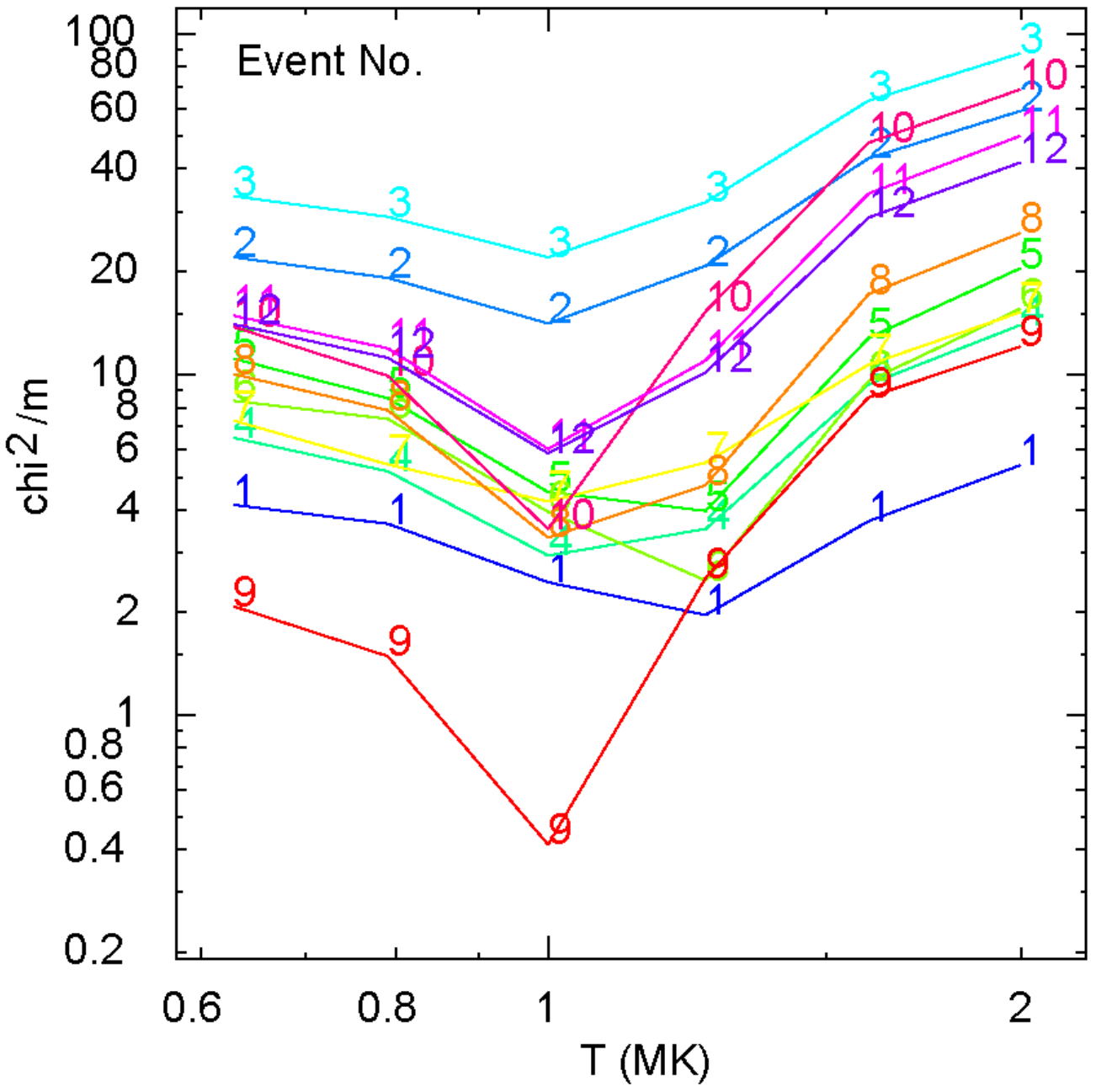

**Figure 4**. Goodness of fit $\chi^2/m$ is shown *vs.* the source or freezing-in temperature $T$ for 12 CIR events. All reach minima at either 1.0 or 1.26 MK. Curves of $A/Q$ *vs.* $T$ that were used are shown by Reames (2016a, 2018b).

Values of $\chi^2/m$ for events 2 and 3 are quite high and excess Fe is seen in the fits especially for event 3 in Figures 3 and 5, perhaps indicating some Fe background from impulsive SEP events.

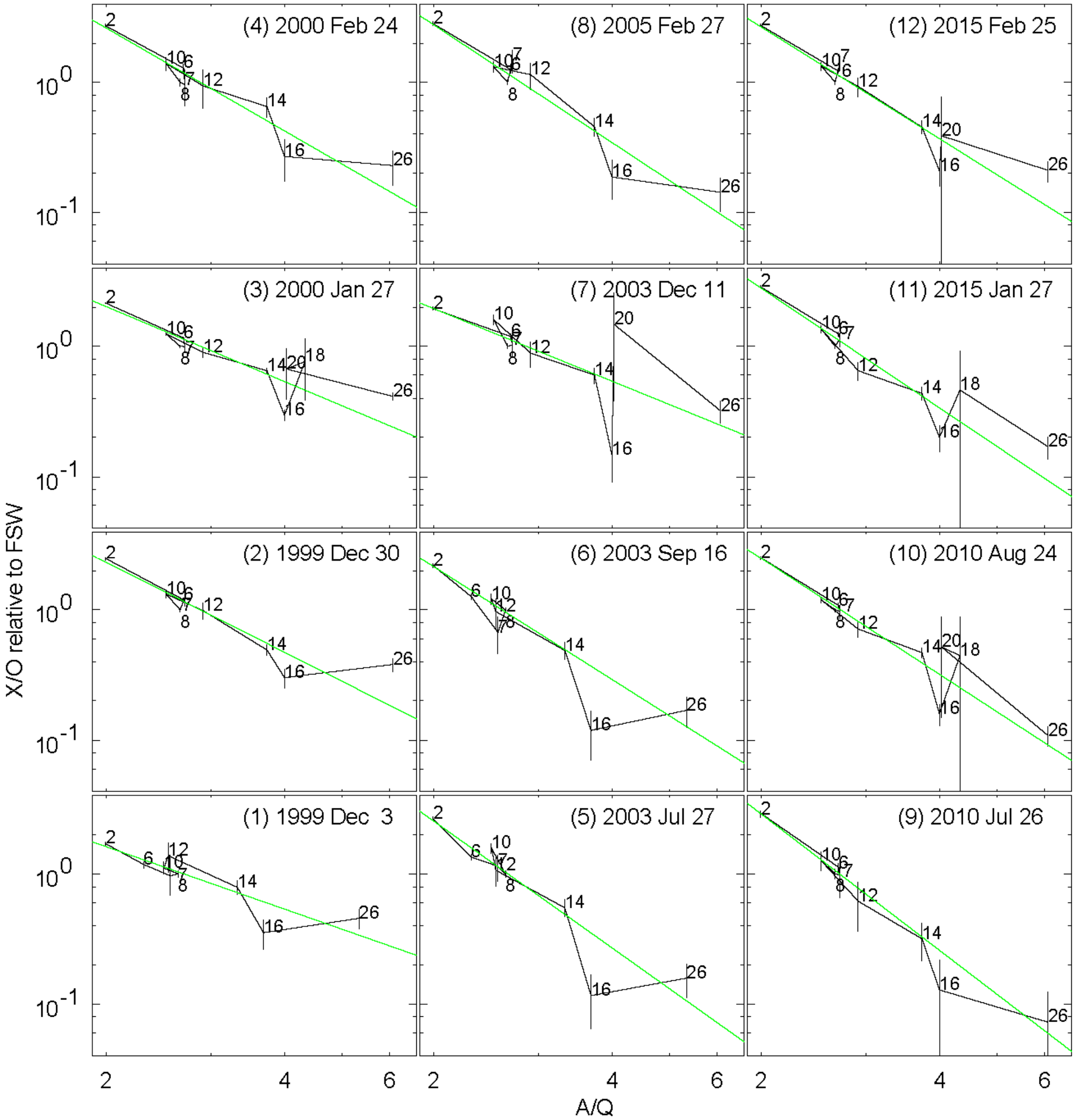

**Figure 5**. Best-fit power laws (green) and element abundances relative to O, divided by corresponding abundances in the FSW (Bochsler, 2009), are plotted *vs. A/Q* for each CIR event. The atomic number *Z* is shown at the position of each element and successive elements are joined. The reference FSW abundance of S (*Z* = 16) seems to be consistently too large in all of the events.

While the power law fits seem appropriate in most cases, the value of S/O= 0.05 in the FSW seems too large by a factor of almost 2 for these CIRs. Also, the value of Fe/O=0.088 could be increased somewhat. These changes would reduce the minimum $\chi^2/m$ for the events. However, it is risky to try to use these fits to redefine the reference abundances for all the elements used to derive them.

In general, we assume that the CIR intensities take the form

$$j(v)=j_0(A/Q)^a v^b \exp(-v/v_0) \tag{1}$$

where $v$ is the ion speed and $j_0$, $a$, $b$, and $v_0$ are adjustable constants. This is the form derived by Fisk and Lee (1980) with explicit dependence on $A/Q$ added. Fisk and Lee (1980) absorb any $A/Q$ dependence into $j_0$. In general, $a \neq b$, and both vary from event to event. Table 1 shows the deduced temperatures and the value of $a$ for the 12 CIR events we have studied using LEMT data.

**Table 1**. CIR Event Intervals, Source *T*, and Power of *A/Q*

| | *Begin* | *End* | *T (MK)* | *a* |
|---|---|---|---|---|
| 1 | 1999 Dec 4 0000 | 1999 Dec 6 0000 | 1.26 | -1.60±0.19 |
| 2 | 1999 Dec 31 0800 | 2000 Jan 4 0000 | 1.00 | -2.30±0.27 |
| 3 | 2000 Jan 28 0000 | 2000 Jan 30 0000 | 1.00 | -1.91±0.20 |
| 4 | 2000 Feb 25 0000 | 2000 Feb 27 0000 | 1.00 | -2.62±0.26 |
| 5 | 2003 Jul 28 0000 | 2003 Aug 1 0000 | 1.26 | -3.24±0.32 |
| 6 | 2003 Sep 17 0000 | 2003 Sep 20 1200 | 1.26 | -2.88±0.24 |
| 7 | 2003 Dec 12 0000 | 2003 Dec 19 0000 | 1.00 | -1.85±0.24 |
| 8 | 2005 Feb 28 0000 | 2005 Mar 14 0000 | 1.00 | -3.00±0.24 |
| 9 | 2010 Jul 27 0000 | 2010 Jul 29 0000 | 1.00 | -3.49±0.16 |
| 10 | 2010 Aug 25 0000 | 2010 Aug 31 0000 | 1.00 | -2.96±0.14 |
| 11 | 2015 Jan 28 0000 | 2015 Jan 30 0000 | 1.00 | -3.06±0.22 |
| 12 | 2015 Feb 26 0000 | 2015 Feb 28 0000 | 1.00 | -2.85±0.23 |
| 13 | 1982 May 2 0000<br>1982 May 29 0000 | 1982 May 10 0000<br>1982 Jun 6 0000 | 1.26 | -0.12±0.21 |

For the gradual SEP events we saw both increasing and decreasing power-law dependence on *A/Q* in a single event (Reames, 2016a). Often, intensities increase with *A/Q* early in an SEP event and decrease later, so we might expect average abundances to recover the source values. For CIRs, abundances are spatial-equilibrium values and need not vary in time in the corotating pattern; there is no "early" or "late" and they always seem to decrease with increasing *A/Q*. Other than consistency of the fits with the FSW abundances in Figure 5, how can we independently determine source abundances for CIRs? What determines $a$? Do we ever find $a = 0$? The CIR abundances from Reames, Richardson, and Barbier (1991) from the CIR event of May 1982 seemed to agree reasonably well with the solar wind abundances of Bochsler (2009) and even support the open-field calculations of Laming (2015) as shown by Reames (2018a, 2018b). In the next section we revisit this 1982 event.

# 5. Revisiting the CIR event of May 1982

Time and energy dependence of element abundances from the ISEE-3 VLET are still available and we can reprocess these data for the recurrent CIR event of May-June 1982 that was first published by Reames, Richardson, and Barbier (1991). Element resolution for the VLET was shown by Reames (1995).

When we sum the intensities over the event periods listed in Table 1, divide by the FSW abundances and plot them *vs. A/Q*, we find the result shown in Figure 6. We use the *A/Q* values for a temperature of 1.26 MK although the data are nearly independent of *A/Q* and do not help us determine a temperature.

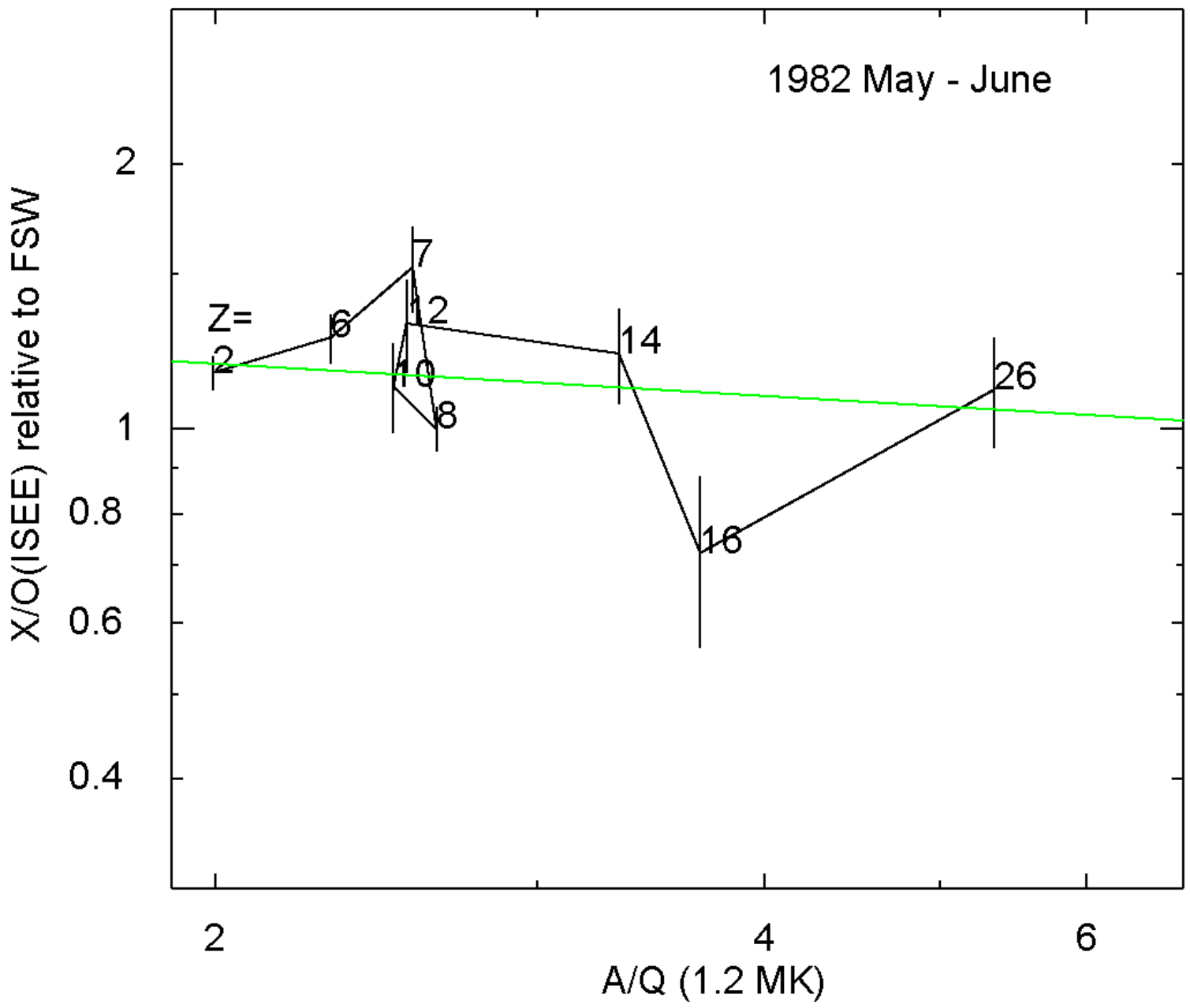

**Figure 6.** Best-fit power law (green) and element abundances relative to O, divided by corresponding abundances in the FSW (Bochsler 2009), are plotted *vs. A/Q* for the CIR event measured seen on ISEE-3 in May – June 1982. The atomic number *Z* is shown at the position of each element and successive elements are joined.

For the CIR event shown Figure 6 we find the power *a* = - 0.12 ± 0.21 as shown in Table 1, *i.e. a* ≈ 0. In this event the energetic-ion abundances in the CIR match those of the FSW.

What is different about this event? Here the reverse shock may actually occur at 1 AU at the spacecraft. A step increase in solar-wind speed and decrease in plasma density expected for a reverse shock is visible, but the shock itself seems to fall in a data gap (Reames, Richardson, and Barbier, 1991). The power *a* is probably zero at the shock and becomes negative and decreases with distance sunward of the shock. It is surely affected by transport where scattering has a dependence upon particle rigidity which varies as *A/Q* when ions are compared at constant velocity (constant Mev $amu^{-1}$). Since ions with higher *A/Q* scatter less they tend to spread more easily farther from the source. This is also seen late in large SEP events (see Fig. 5.10 of Reames 2017a). For SEPs, one can see the time dependence, but CIRs involve a time equilibrium.

Since the revised study of the 1982 CIR includes a somewhat increased time period, we compare the new abundances from this event with those from other sources in Table 2. The table shows the photospheric abundances of Caffau *et al.* (2011) supplemented by the meteoric values of Lodders, Palme, and Gail (2009); these differ somewhat from the abundances of Asplund *et al.* (2009). The CIR measurement shown in the table is based upon the re-measured CIR event of 1982 which lacks a dependence upon *A/Q*. Including the other 12 events would require uncertain corrections for their *A/Q* dependence.

**Table 2** Element Abundances from Various Sources

| | Z | FIP [eV] | Photosphere[1] | SEPs[2] | CIRs This Work | Interstream SSW[3] | Coronal Hole FSW[3] |
|---|---|---|---|---|---|---|---|
| H | 1 | 13.6 | $1.74\times10^{6}$ * | $(\approx1.6\pm0.2)\times10^{6}$ | – | – | – |
| He | 2 | 24.6 | $1.6\times10^{5}$ | 91000±5000 | 86700±3700 | 90000±30000 | 75000±20000 |
| C | 6 | 11.3 | 550±76* | 420±10 | 860±54 | 680±70 | 680±70 |
| N | 7 | 14.5 | 126±35* | 128±8 | 175±29 | 78±5 | 114±21 |
| O | 8 | 13.6 | 1000±161* | 1000±10 | 1000±59 | 1000 | 1000 |
| Ne | 10 | 21.6 | 210 | 157±10 | 157±18 | 140±30 | 140±30 |
| Na | 11 | 5.1 | 3.68 | 10.4±1.1 | – | 9.0±1.5 | 5.1±1.4 |
| Mg | 12 | 7.6 | 65.6 | 178±4 | 140±17 | 147±50 | 106±50 |
| Al | 13 | 6.0 | 5.39 | 15.7±1.6 | – | 11.9±3 | 8.1±0.4 |
| Si | 14 | 8.2 | 63.7 | 151±4 | 123±16 | 140±50 | 101±40 |
| P | 15 | 10.5 | 0.529±0.046* | 0.65±0.17 | – | 1.4±0.4 | – |
| S | 16 | 10.4 | 25.1±2.9* | 25±2 | 36±8 | 50±15 | 50±15 |
| Cl | 17 | 13.0 | 0.329 | 0.24±0.1 | – | – | – |
| Ar | 18 | 15.8 | 5.9 | 4.3±0.4 | – | 3.1±0.8 | 3.1±0.4 |
| K | 19 | 4.3 | 0.224±0.046* | 0.55±0.15 | – | – | – |
| Ca | 20 | 6.1 | 3.85 | 11±1 | – | 8.1±1.5 | 5.3±1.0 |
| Ti | 22 | 6.8 | 0.157 | 0.34±0.1 | – | – | – |
| Cr | 24 | 6.8 | 0.834 | 2.1±0.3 | – | 2.0±0.3 | 1.5±0.3 |
| Fe | 26 | 7.9 | 57.6±8.0* | 131±6 | 98±14 | 122±50 | 88±50 |
| Ni | 28 | 7.6 | 3.12 | 6.4±0.6 | – | 6.5±2.5 | – |
| Zn | 30 | 9.4 | 0.083 | 0.11±0.04 | – | – | – |

[1]Lodders, Palme, and Gail (2009).

* Caffau *et al.* (2011).

[2] Reames (1995, 2014, 2017a).

[3] Bochsler (2009).

Figure 7 shows the FIP patterns of the new CIR and the FSW measurements, relative to the photospheric abundances, all taken from Table 2, compared with the open field theoretical result listed in the column labeled 6.0 km$^{-1}$ s in Table 4 of Laming (2015). We have chosen to normalize the data at Ne, rather than O, recognizing that O may actually be becoming a transition element rather than a pure high-FIP element. This normalization seems quite appropriate for Mg, Si, and Fe as well.

**Figure 7**. Element abundances relative to the photosphere *vs.* FIP from Table 2 are compared with the open-field Alfvén-wave theory of Laming (2015). The abundances are normalized at Ne, not O.

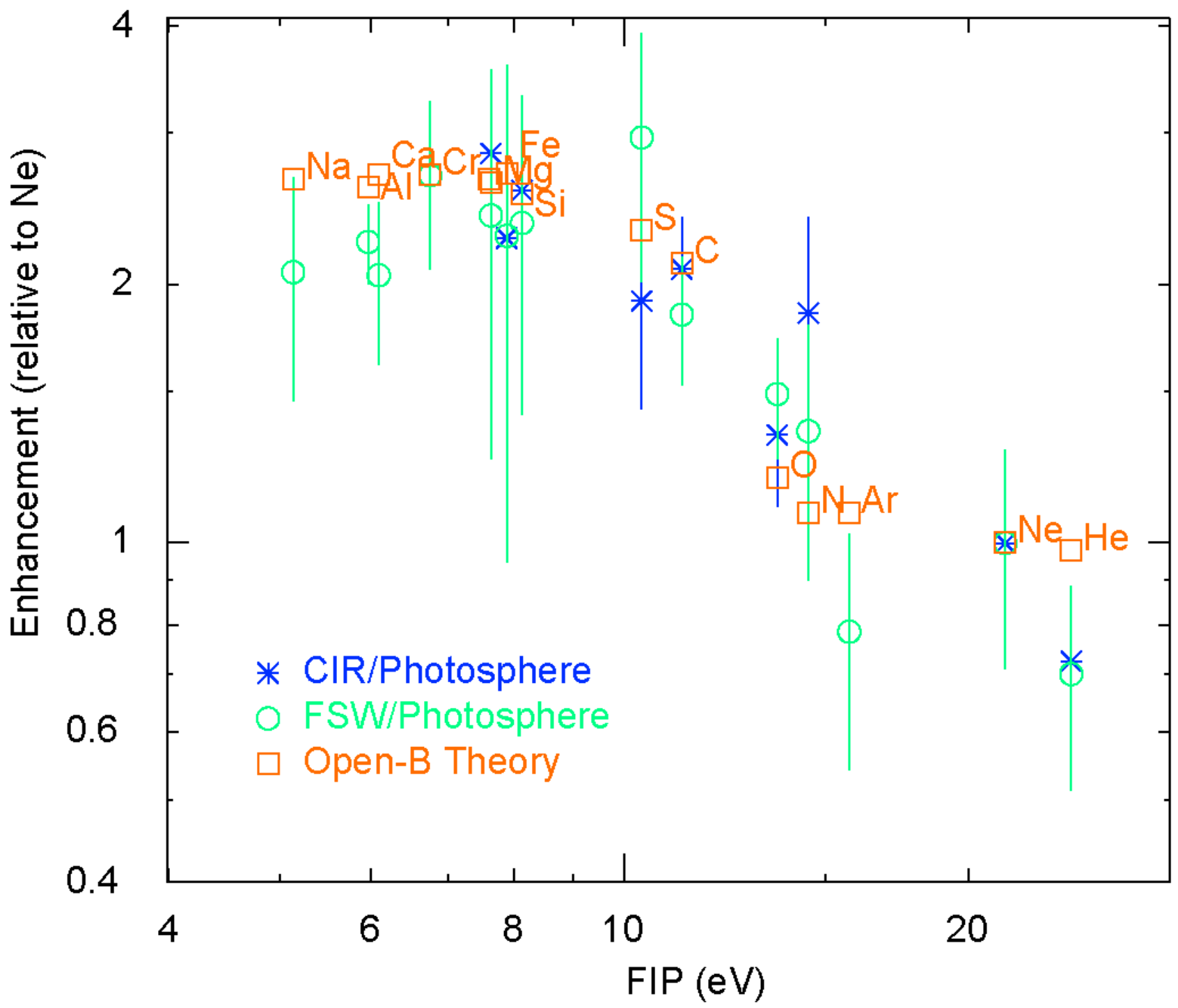

It has been clear in all our CIR measurements that S/O is smaller in CIRs than in the reference solar wind. However, the new CIR value of S/O = 0.036 ± 0.008 is still high enough to suggest than S is close to behaving as a low-FIP transition element similar to C, as predicted by the open-field theory. C, S, and P are the elements that best distinguish the open- and closed-field models; P is not available in CIRs or the FSW, but C is well measured and strongly enhanced in both.

## 6. Discussion

Thus the element abundances of the energetic ions from CIRs, relative to the FSW, show a power-law dependence on *A/Q* of the ions that is similar to that seen late in SEP events. Fisk and Lee (1980) discussed the velocity dependence of the spectra and their depend-ence on distance from the shock. They showed that spectra of different ions were similar in *v*, but they did not examine the normalization of the abundances or their dependence on *A/Q*. Measurements of spectral evolution with time (*e.g.* Reames *et al.*, 1997) may only involve a single species; they include different source locations along the shock as well as different distances from it. Unfortunately, we cannot directly measure distances from a remote shock wave.

In general, multi-spacecraft measurements (e.g. Van Hollebeke, McDonald and von Rosenvinge, 1978; Richardson. 2004) show that CIR shocks are *much* stronger in the outer heliosphere. The fast wind flows parallel to the slow wind near the Sun, but begins to “bite” into it more and more as *R* increases. Shocks that accelerate ions usually form beyond 1 AU. CIR shocks are much stronger and ion intensites are much higher in the outer heliosphere at ~5 AU (Van Hollebeke, McDonald, and von Rosenvinge, 1976) yet the Fisk and Lee (1980) formula continues to fit the 1 AU He spectrum for up to 8 days later (Reames *et al.,* 1997). Intensities at 1 AU must be a balance between shock acceleration and transport, including possible field-line meandering. He ions at high energies (*i.e.* high rigidities, which scatter less) have been observed to peak many days later than those at low energy (see Reames *et al.*, 1997).

The CIR event of May – June 1982 is unusual in having a reverse shock with $a \approx 0$ that seems to form near 1 AU. The parameter *a* must depend upon transport since scattering depends upon particle rigidity and hence upon *A/Q*. None of the CIR events we have measured in this paper from the *Wind* era from November 1994 to June 2018 have the property $a \approx 0$. There are just 9 fast reverse shock waves in the database of shocks, seen at 1 AU, with complete analysis of the shock parameters, measured on the *Wind* spacecraft by J. Kasper (http://www.cfa.harvard.edu/shocks/wi_data/, see also Jian *et al.*, 2006). None of them provide energetic (2 – 10 MeV $amu^{-1}$) ions for us to study.

The suprathermal seed population in CIRs was studied by Chotoo *et al.* (2000). They found that ≈15% of the He was singly ionized He from pickup ions. Corrections for other pickup ions are probably negligible. Suppression of ions with high *A/Q* in events with distant shocks would strongly suppress any singly-ionized pickup ions.

An alternative explanation for the high C/O value in CIR events was advanced by Gloeckler *et al.* (2000) to be an ‘inner source’ where interstellar grains capture and re-lease solar wind ions to produce $C^+$ and $O^+$. However, such ions would be strongly suppressed in most CIR events because of their high values of *A/Q* with *Q* = 1. The FSW seems a much more likely source of energetic ions in CIRs, although pre-energized SEPs could contribute.

Previous attempts have been made to study *A/Q* dependence in CIR events (*e.g.* Richardson *et al.*, 1993). Any successful study of *A/Q* dependence requires high-quality reference abundances (*e.g.* Bochsler, 2009) and a plausible regime of source temperatures to define *A/Q* values. These have only recently come together. There are many CIR observations that seem to have enhanced He/O and suppressed Fe/O (*e.g.* Mason *et al.*, 2008; Bučík *et al.*, 2012). However, presence of SEP background, especially impulsive SEP events, could enhance Fe/O and discourage a search for *A/Q* dependence in CIRs.

Although actually predicted by theory (Parker, 1963), the power-law dependence of SEP abundances on *A/Q*, reported by Breneman and Stone (1985), was a significant finding. Unfortunately, it languished for decades, perhaps because it took time to recognize that *Q* and hence *A/Q* have an important dependence upon source temperature, which can vary among SEP events. Finally it seems clear that this power law behavior extends to energetic ions from CIRs as well as SEPs and is a general property of energetic-particle scattering and transport in space. We would encourage a theoretical study the *A/Q* dependence in CIRs.

## Disclosure of Potential Conflicts of Interest

The author declares he has no conflicts of interest.